\documentclass{article}

\usepackage[final]{nips_2017}
\usepackage{graphicx}
\usepackage{multirow}
\usepackage{url}

\usepackage[utf8]{inputenc} 
\usepackage[T1]{fontenc}    
\usepackage{hyperref}       
\usepackage{url}            
\usepackage{booktabs}       
\usepackage{amsfonts}       
\usepackage{nicefrac}       
\usepackage{microtype}      

\title{FlagIt: A System for Minimally Supervised Human Trafficking Indicator Mining}

\author{
 Mayank Kejriwal, Jiayuan Ding, Runqi Shao, Anoop Kumar and Pedro Szekely \\
  Information Sciences Institute, USC Viterbi School of Engineering\\
  4676 Admiralty Way, Ste. 1001\\
  Marina Del Rey, CA 90292 \\
  \texttt{\{kejriwal,jiayuand,runqisha,anoopk,pszekely\}@isi.edu} \\
}

\begin{document}

\maketitle

\begin{abstract}
 In this paper, we describe and study the indicator mining problem in the online sex advertising domain. We present an in-development system, FlagIt (Flexible and adaptive generation of Indicators from text), which combines the benefits of both a lightweight expert system and classical semi-supervision (heuristic re-labeling) with recently released state-of-the-art unsupervised text embeddings to tag millions of sentences with indicators that are highly correlated with human trafficking. The FlagIt technology stack is open source. On preliminary evaluations involving five indicators, FlagIt illustrates promising performance compared to several alternatives. The system is being actively developed, refined and integrated into a domain-specific search system used by over 200 law enforcement agencies to combat human trafficking, and is being aggressively extended to mine at least six more indicators with minimal programming effort. FlagIt is a good example of a system that operates in limited label settings, and that requires creative combinations of established machine learning techniques to produce outputs that could be used by real-world non-technical analysts.   
\end{abstract}

\section{Introduction}

The growth, combined with the ease of sharing information on the Web, has also led to increased illicit activity both on the Open and Dark Web, an egregious example being human trafficking (HT) \cite{ht1}, \cite{ht2}. The DARPA MEMEX program, which funds research into domain-specific search, has collected hundreds of millions of online sex advertisements, a significant (but unknown) number of which are believed to be sex (and human) \emph{trafficking} instances\footnote{\url{https://www.darpa.mil/program/memex}}.    


Flagging scraped content from webpages in such domains with activity tags or \emph{indicators}, denoted herein as the \emph{indicator mining} problem, has an investigative and socially motivated purpose not just in online human trafficking but several other domains that have also been studied in MEMEX, including patent trolling, counterfeit electronic sales, illegal weapons sales and narcotics\footnote{The authors have personally encountered a variant of indicator mining in each of these domains, although this paper is limited to human trafficking.}. In recent years, some of these `illicit' Web domains have been intensely studied by social and computer scientists seeking to profile activity on the Dark Web \cite{darkweb1}, \cite{darkweb2}; however, to the best of our knowledge, flagging Web content with tags indicating illicit activity has received considerable less attention.

An indicator is a flag, typically binary, but potentially multi-categorical, that is suggestive of suspicious activity that would warrant investing more resources into investigating the subject of the content being flagged. 
In the case of the online sex advertisement domain, subjects are not only escorts (usually, the victims of trafficking) but also perpetrator organizations like massage parlors, procurers, as well as reviewers and clients. While the ideal motivation is to directly flag an online sex trafficking document with a single human trafficking indicator, this is extremely problematic in practice, especially without a proper field level investigation. A more attainable goal in data mining is to instead use the mined indicators to \emph{guide} further investigation. To distinguish trafficked escorts from non-trafficked escorts, investigators seek signals that suggest (whether at present or in the future), for example, that the subject exhibits \emph{movement} between cities or provides \emph{risky} services that were highly correlated with trafficking activity in the past. Another example of a tag studied in this paper, \emph{multi-girl}, flags webpage content that is suggestive of simultaneous advertising of multiple women. Content tagged in this way becomes a fruitful target, both for investigators and non-governmental organizations, for implementing preventive and prosecution measures. 

Computationally, this task is challenging for various reasons, the most prominent being the unusual nature of the domain, the lack of training data and multiple forms of bias that emerge in the corpus due to the nature of crawling systems tuned for portals with significant presence of sex ads and reviews. Some of these problems were studied independently in both \cite{benpaper} and \cite{ht1}. 

In online sex advertisements, the text content being tagged is typically extracted from the main body of an advertisement scraped from a Web domain that is known (by domain experts building the relevance model for the crawling) to have high prevalence of online sex trafficking activity. An example of such a Web domain is the adult section of \url{backpage.com}, which contains many ads, both for escorts as well as brick-and-mortar operations like massage parlors that are fronting illicit sex trafficking. 

\begin{figure*}
\centering
\includegraphics[height=2.1in, width=5.7in]{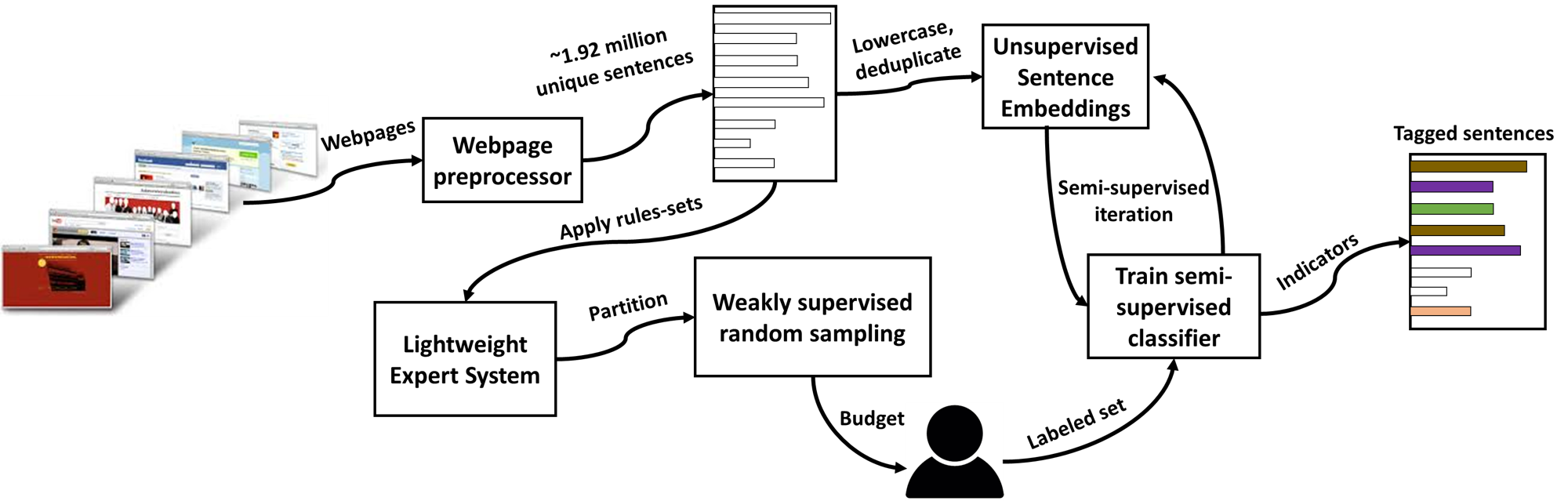}
\caption{A high-level overview of FlagIt, which accepts a (possibly streaming) domain-specific corpus of crawled webpages as input, extracts sentences from those pages and tags them with a set of indicators. The output of FlagIt is deployed in a GUI (not shown) and can be used to quickly zoom in on suspicious pages and text within the corpus. }\label{fig:approach}
\end{figure*}

In this paper, we present the architecture (Figure \ref{fig:approach}) of an end-to-end indicator mining approach called FlagIt ({\bf Fl}exible and {\bf a}daptive {\bf g}eneration of {\bf I}ndicators from {\bf t}ext). To reduce the burdens of manual supervision and potential user bias, FlagIt combines a lightweight expert system with applied research (both recent and classic) in minimally supervised machine learning, including unsupervised text embeddings and semi-supervised heuristic re-labeling \cite{semisup}, to achieve average F-Measure scores slightly below 80\%. Also, compared to four baselines, both adaptive and non-adaptive, FlagIt outperforms the best baseline by 2-13\% F-Measure on four of the five indicators and is always the best of all adaptive baselines. FlagIt is simple and extensible, and scales to millions of sentences.

Indicator tags discovered by FlagIt are designed to be used both for lead generation and lead investigation. Lead generation arises when investigators search from scratch e.g., an investigator may be interested in investigating the illegal operations of brothels that are fronting as legitimate massage parlors on paper (or by day). Lead investigation is a more focused search, often involving the gathering of evidence, where some information e.g., a phone number, is already available and suspected of involving trafficking activity. Using indicators produced by FlagIt, and the exploratory search capabilities produced as a result of other research in MEMEX, investigators can hone in on crucial evidence without conducting expensive fieldwork or issuing time-consuming subpoenas.       

\section{Related Work}\label{rw}
FlagIt draws on a combination of established techniques in limited-label machine learning, knowledge base construction \cite{deepdive}, text embeddings \cite{fasttext}, \cite{docvec} and expert systems \cite{expertsurvey1}, as well as recent conceptual advances that have pushed the state-of-the-art \cite{snorkel}. We provide details on relevant sub-topics below. An important contribution of FlagIt is to test the practical \emph{application} and \emph{combination} of established techniques in an unusual domain like human trafficking. To the best of our knowledge, successful data mining results in the HT domain have been very limited, mostly restricted to extensively studied areas such as information extraction \cite{kejriwalgeorich} and generic document clustering \cite{benpaper}. However, interest in the area has been increasing, a recently published example being \cite{semisupHT}.

\section{Architecture}\label{approach}
FlagIt assumes a fixed corpus of webpages. We construct the sentence corpus from Web resources that are known to contain high levels of human trafficking (HT) activity. The sentences are obtained by executing a webpage pre-preprocessing pipeline that was tuned to scale and generalize to a 54 GB HT corpus containing webpages from numerous \emph{Web domains}. First, we use the Readability Text Extractor\footnote{RTE was available and open-source at the time of preprocessing. It has since been taken offline and superseded by a new web scraping API called Mercury: \url{https://www.readability.com/}} (RTE) to scrape descriptive content from each webpage in the corpus. We tuned RTE for high recall i.e. with high probability, all relevant sentences constituting the main content of the webpage are extracted, but some irrelevant sentences (e.g., `taylor profile was posted 14 weeks 4 days ago') may also be extracted. Next, the output of RTE is segmented into a list of sentences by using newlines and consecutive occurrences of three whitespace characters as delimiters. All sentences were converted to lowercase and deduplicated before executing the lightweight expert system (LES), which relies on capitalizations and other text features to perform well. Following LES execution, the minimally supervised machine learning modules (e.g., the text embeddings) are executed.

We briefly describe some of the critical aspects of the LES, as it is an important component in FlagIt. At its core, the LES relies on shallow \emph{pattern matching} rules, with each rule exclusively falling into one of four \emph{rule categories} depending on its predictive relationship to the indicator: positive (P), strong positive (Sp), negative (N) and strong negative (Sn). The `strong' qualifier is meant to express user confidence, providing users a simple but effective way of expressing when they are very sure about a rule, and when they are not. The categories are motivated by extensive research in cognitive science and crowdsourcing showing that asking users for fine-grained confidence outputs (including probabilities) is rife with issues of consistency and labeling context (a notorious example being the order in which outputs are elicited).

We define a pattern matching rule as a finite sequence of \emph{pattern elements}, where a pattern element can either be a \emph{constant} token (a finite sequence of Unicode characters) or a pattern variable $V$. In keeping with the lightweight nature of the system, four kinds of pattern variables are considered. We describe these variables below, followed by some representative examples.

\emph{Glossary Variable.} Given a glossary $G$ (a finite set of tokens), a glossary variable $V_G$ can take exactly one token from $G$ as its value. In some cases, glossaries are easily available from the Web (e.g., a list of common English names) but in other cases, have to be specified either by the expert or through entity set expansion and glossary mining \cite{ese}. For the indicators considered in this paper, five glossaries are used for both \emph{incall} and \emph{outcall}, two glossaries each for \emph{movement} and \emph{risky}, and four glossaries for \emph{multi-girl}.

\emph{Regular Expression (RegEx).} Pattern variables can also be encoded as regular expressions. For efficiency (and also noting that domain experts are non-technical), we limit expressivity of regular expressions in FlagIt. The most commonly used RegEx was to check for tokens possibly symbolizing names and geolocations by specifying capitalization and alphabetic constraints on the tokens.

\emph{NLP Tags.} Despite being lightweight, the expert system can make use of tags, such as POS tags and English dependency labels\footnote{See \url{https://spacy.io/docs/api/annotation} for a description.}, output by NLP packages like spaCy \cite{spacy}. As a subsequent example illustrates for \emph{incall}, dependency labels, such as an \emph{adjectival modifier} tag, can be used to great effect without complex overhead.

\emph{Named Entities.} The LES can also use an expected named entity \emph{type} as a pattern  variable specification. The variable is assigned the word, only if it is extracted as a named entity of that type \cite{ner}. 

\emph{Example.} Given a glossary \emph{place}, containing residence terms and synonyms such as \emph{apartment} and \emph{studio}, and another glossary \emph{priv} containing words such as \emph{private} and \emph{discreet}, a simple but effective Sp rule for \emph{incall} is the sequence $(G_{priv}<-amod, G_{place})$, which would match a pattern such as \emph{private apartment} in the text. The $<-amod$ specifies that the word must have an adjectival modifier dependency tag. This example shows that finite combinations of pattern variables are also possible. In a slight abuse of notation, we use the symbols P, Sp, N and Sn to respectively refer to the collections of pattern matching rules defined by domain experts. A rule $R$, when applied to a sentence, yields \emph{True} if at least one pattern is successfully matched in the sentence (said to be \emph{covered} by the rule). 

Because rules from multiple rule categories can cover a sentence, the notion of a \emph{rules-set} becomes necessary to enforce a \emph{partition} of the corpus. The FlagIt LES supports seven rules-sets, resulting in a 7-split partition: P OR Sp, N OR Sn, Sp AND N, Sn AND P, Sp AND Sn, P AND N, and Null. The last is when no rule (from any of the four categories) yields \emph{True}. OR in a rule such as P OR Sp indicates that only a rule from P or Sp (or both) covers a sentence. AND in a rule such as Sp AND N indicates that the sentence must be covered by a rule from Sp, as well as from N. Where applicable, Sn takes \emph{precedence} over N and Sp over P. For example, if a rules-set such as (P AND Sp) AND Sn  fires, it is considered equivalent to the rules-set Sp AND SN firing.

Because of the specified precedence, the rules-sets described above are such that a sentence will be covered by exactly one of them, yielding a partition over a sentence corpus. This partition is used in a weakly supervised setting to obtain labeled data that can then be used in a minimally supervised machine learning setting.

\subsection{Minimally Supervised Components}\label{tsc}


Given a sentence from the corpus of segmented sentences, the indicator mining problem can be modeled as \emph{multi-label}\footnote{Distinguished herein from \emph{multi-class}, since a sentence could be labeled with multiple indicators.} text classification, assuming that representative training data and feature functions are available. Unfortunately, we have no training data available for this task, and enormous \emph{negative} class skew (higher than 90\% in many cases) in the task domain preempts random sampling and labeling. Because indicator mining is an emerging problem in data mining, it is also not clear what indicator-specific features will lead to good performance for a given indicator. FlagIt attempts to address these challenges in a simple and lightweight manner. 

The version of FlagIt described in this paper implements five indicator classifiers, namely \emph{incall}, \emph{outcall}, \emph{movement}, \emph{risky} and \emph{multi-girl}. \emph{Incall} explicitly indicates that a client must visit an escort (conversely for \emph{outcall}) at a specified location. \emph{Movement} indicates that the escort is visiting the locale (usually for short periods) from a different previous locale, while \emph{risky} indicates that the escort is willing to engage in risky sexual activities (as determined by domain experts). \emph{Multi-girl} (along with \emph{movement}) is highly correlated with organized sex trafficking activity\footnote{In an extended version of FlagIt, an explicit indicator for \emph{massage parlor} activity is also considered. This indicator can overlap with, but is technically distinct from, \emph{multi-girl}.}, and indicates that multiple girls are being simultaneously advertised.      

FlagIt follows a \emph{weakly supervised random sampling} scheme for training set construction.  First, we construct a small but diverse training set per indicator by randomly sampling a small number of sentences from each of the seven rules-sets in the partition. In keeping with the minimal supervision goals of this paper, \emph{labeling budget} was restricted to a small number (140), meaning that, in the typical case, 20 sentences (for each rules-set category) were labeled by a user. The only exception is when a rules-set did not cover 20 sentences, in which case we used a form of iterative, re-distributive sampling whereby each sentence covered by that rules-set was labeled, and the rest of the budget was divided between the remaining rules-sets that had more than 20 sentences\footnote{The process is iterative because a rules-set may not now cover the \emph{new} local labeling budget (by definition, greater than 20).}. In the worst case, the scheme would be reduced to sampling all 140 sentences from the Null rules-set but this never happened in practice for any of the five indicators in FlagIt. 





Two options to considerably mitigate or even eliminate the twin labors of labeling and feature crafting effort are semi-supervised learning and unsupervised text embeddings respectively. While the former has been long studied, the latter is an intense topic of recent research. FlagIt leverages both.
First, given that the weakly supervised random sampling scheme described earlier for constructing the evaluation set for each indicator is extremely small compared to the whole corpus, we posit that semi-supervised learning can be used to leverage this body of unlabeled text to further improve the results \cite{semisup}. We integrate a simple, and classic semi-supervised scheme in FlagIt: first, we train an initial (indicator-specific) classifier on the perfectly labeled training set, and use the classifier to generate a positive label \emph{probability} (for that indicator) for each unlabeled sentence in the overall corpus. Next, we pick some percentage of unlabeled sentences from the `extreme' ends of the probability distribution and heuristically label these sentences. We re-train the classifier by supplementing the perfectly labeled training set with the heuristically labeled sentences. This process can be iterated a certain constant number of steps, or till some convergence criterion is met. The final classifier is tested on a held-out test set. We found in early experimentation that even a relatively simple scheme of heuristically relabeling 45\% (from each end of the probability distribution) unlabeled data, and a single iteration, yields significant improvements over not using any form of semi-supervised learning.  

\section{Brief Summary of Experimental Results}\label{experiments}

We evaluated\footnote{Because of the sampling scheme, classes were relatively balanced during both training and testing.} FlagIt on five diverse indicators against a range of competitive baselines, with encouraging results. Compared to four baselines, three adaptive\footnote{Resp. based on bag-of-words, limited label supervised fasttext and paragraph2vec \cite{docvec}.} and one non-adaptive\footnote{A highly tuned version of the LES component.}, FlagIt outperforms the best baseline on the F1-Measure metric by 2-13\% on four of the five indicators and is always the best of all adaptive baselines. We also found that, while semi-supervised learning always improves performance in FlagIt (by 1-9\%), the choice of the unsupervised text embedding algorithm can be crucial. For example, the recently released fasttext package \cite{fasttext}, based on a \emph{bag-of-tricks} approach, proved to be a much more suitable fit for an irregular domain like human trafficking than the widely used paragraph2vec algorithm \cite{docvec}, which performed badly despite extensive parameter tuning.

\section{Impact, Deployment and Future Work}\label{impact}

FlagIt is already being integrated into one of two end-to-end domain-specific search engines (denoted herein as DS1 and DS2 to maintain anonymity) being developed under the DARPA MEMEX program, with possible integration into the other system (DS2) in the medium term. The key idea is to show webpages highly ranked with respect to a given indicator, either because of highly indicative sentences, or the number of sentences with non-trivial indicator probability. 
Both DS1 and DS2 are currently being used by over 200 law enforcement agencies in the US to combat human trafficking. The tools are complementary to each other. Indicator mining is an important application of both search and lead generation, especially considering the limited resources of law enforcement. Analysis is showing that indicator co-occurrences show unusual correlations with each other, which raises interesting avenues both for quantitative social science research, and for complex multi-label classification models. Data collection is currently ongoing to quantify such correlations.  Also, although only five indicators have been formally evaluated thus far, FlagIt is designed to support between 11-13 (and potentially many more) indicators specified by domain experts at the time of writing. Because of its lightweight, minimally supervised architecture, this extension is possible with only a few hours of effort, divided between specifying rules in spaCy, sampling sentences, and providing manual indicator labels. Another important line is research is supporting FlagIt for explainable machine learning, for which we are investigating two options: highlighting highly indicative sentences flagged by FlagIt in the GUI, or using the approach espoused by \cite{sandholm} with the pattern matching rules in the lightweight expert system serving the role of locally explainable classifiers.

\bibliographystyle{plain}
\bibliography{sigproc} 

\begin{thebibliography}{10}

\bibitem{spacy}
spacy nlp toolkit.
\newblock https://spacy.io/docs/usage/resources, 2017.
\newblock Accessed: 2017-08-12.

\bibitem{semisupHT}
Hamidreza Alvari, Paulo Shakarian, and JE~Kelly Snyder.
\newblock Semi-supervised learning for detecting human trafficking.
\newblock {\em Security Informatics}, 6(1):1, 2017.

\bibitem{docvec}
Andrew~M Dai, Christopher Olah, and Quoc~V Le.
\newblock Document embedding with paragraph vectors.
\newblock {\em arXiv preprint arXiv:1507.07998}, 2015.

\bibitem{benpaper}
Artur Dubrawski, Kyle Miller, Matthew Barnes, Benedikt Boecking, and Emily
  Kennedy.
\newblock Leveraging publicly available data to discern patterns of
  human-trafficking activity.
\newblock {\em Journal of Human Trafficking}, 1(1):65--85, 2015.

\bibitem{fasttext}
Armand Joulin, Edouard Grave, Piotr Bojanowski, and Tomas Mikolov.
\newblock Bag of tricks for efficient text classification.
\newblock {\em arXiv preprint arXiv:1607.01759}, 2016.

\bibitem{kejriwalgeorich}
Rahul Kapoor, Mayank Kejriwal, and Pedro Szekely.
\newblock Using contexts and constraints for improved geotagging of human
  trafficking webpages.
\newblock In {\em Proceedings of the Fourth International ACM Workshop on
  Managing and Mining Enriched Geo-Spatial Data}, page~3. ACM, 2017.

\bibitem{ht1}
Mark Latonero.
\newblock Human trafficking online: The role of social networking sites and
  online classifieds.
\newblock 2011.

\bibitem{expertsurvey1}
Shu-Hsien Liao.
\newblock Expert system methodologies and applications - a decade review from
  1995 to 2004.
\newblock {\em Expert systems with applications}, 28(1):93--103, 2005.

\bibitem{ner}
David Nadeau and Satoshi Sekine.
\newblock A survey of named entity recognition and classification.
\newblock {\em Lingvisticae Investigationes}, 30(1):3--26, 2007.

\bibitem{deepdive}
Feng Niu, Ce~Zhang, Christopher R{\'e}, and Jude~W Shavlik.
\newblock Deepdive: Web-scale knowledge-base construction using statistical
  learning and inference.
\newblock {\em VLDS}, 12:25--28, 2012.

\bibitem{ese}
Patrick Pantel, Eric Crestan, Arkady Borkovsky, Ana-Maria Popescu, and Vishnu
  Vyas.
\newblock Web-scale distributional similarity and entity set expansion.
\newblock In {\em Proceedings of the 2009 Conference on Empirical Methods in
  Natural Language Processing: Volume 2-Volume 2}, pages 938--947. Association
  for Computational Linguistics, 2009.

\bibitem{snorkel}
Alexander~J Ratner, Christopher~M De~Sa, Sen Wu, Daniel Selsam, and Christopher
  R{\'e}.
\newblock Data programming: Creating large training sets, quickly.
\newblock In {\em Advances in Neural Information Processing Systems}, pages
  3567--3575, 2016.

\bibitem{sandholm}
Marco~Tulio Ribeiro, Sameer Singh, and Carlos Guestrin.
\newblock Model-agnostic interpretability of machine learning.
\newblock {\em arXiv preprint arXiv:1606.05386}, 2016.

\bibitem{ht2}
Elizabeth~M Wheaton, Edward~J Schauer, and Thomas~V Galli.
\newblock Economics of human trafficking.
\newblock {\em International Migration}, 48(4):114--141, 2010.

\bibitem{darkweb1}
Jennifer Xu and Hsinchun Chen.
\newblock The topology of dark networks.
\newblock {\em Communications of the ACM}, 51(10):58--65, 2008.

\bibitem{semisup}
Xiaojin Zhu.
\newblock Semi-supervised learning literature survey.
\newblock 2005.

\bibitem{darkweb2}
Ahmed~T Zulkarnine, Richard Frank, Bryan Monk, Julianna Mitchell, and Garth
  Davies.
\newblock Surfacing collaborated networks in dark web to find illicit and
  criminal content.
\newblock In {\em Intelligence and Security Informatics (ISI), 2016 IEEE
  Conference on}, pages 109--114. IEEE, 2016.

\end{thebibliography}

\end{document}